\documentstyle[psfig]{mn-nat}
\begin{document}
\label{firstpage}
\title{\bf GMRT observations of four suspected supernova remnants near the
Galactic Centre}
\author[Subhashis Roy and A. Pramesh Rao]
{Subhashis Roy and A. Pramesh Rao \\
National Centre for Radio Astrophysics (TIFR), \\
Pune University Campus, Post Bag No.3, Ganeshkhind, Pune 411 007,
India.\\ 
E-mail: roy@ncra.tifr.res.in, pramesh@ncra.tifr.res.in }

\date{}
\pagerange{\pageref{firstpage}--\pageref{lastpage}}

\maketitle
\begin{abstract}
We have observed two fields - Field I (l=3.2$^{\circ}$,
b=$-$1.0$^{\circ}$) and Field II (l=356.8$^{\circ}$,
b=$-$0.1$^{\circ}$) with the Giant Metrewave Radio
Telescope (GMRT) at 330 MHz. In the first field, we
have studied the candidate supernova remnant (SNR)
G3.1$-$0.6 and based on its observed morphology,
spectral index and polarisation confirmed it to be an
SNR. We find this supernova to have a double ring
appearance with a strip of emission on it's western
side passing through it's centre. 

We have discovered two extended curved objects in the second
field, which appears to be part of a large shell like
structure. It is possibly the remains of an old supernova in the
region. Three suspected supernova remnants, G356.3$-$0.3,
G356.6+0.1 and G357.1$-$0.2 detected in the MOST 843 MHz
survey of the Galactic Centre region appears to be located
on this shell like structure. While both G356.3$-$0.3 and
G356.6+0.1 seem to be parts of this shell, G357.1$-$0.2
which has a steeper spectrum above 1 GHz, could be a
background SNR seen through the region.  Our HI absorption
observation towards the candidate SNR G357.1$-$0.2 indicates
that it is at a distance of more than 6 kpc from us.

\end{abstract}


\section{Introduction:}
Since present catalogues of SNRs are thought to be
incomplete due to different sensitivities and angular
resolutions of various telescopes used for surveys,
the distribution of SNRs in our Galaxy is not well
constrained. This problem is further compounded in the
Galactic Centre (GC) region, where, there is a very
complex distribution of sources.  13 SNRs had been
identified within 355$^{\circ} \ge l \le 5^{\circ}$,
$-2.5^{\circ} \ge b \le 2.5^{\circ}$ region of the
Galaxy before the MOST Galactic Centre survey (MGCS).
The MGCS \nocite{1994MNRAS.270..847G}({Gray} 1994b) detected
another 17 candidate SNRs in the region. These detections,
if confirmed, indicate the number density of SNRs in the GC
region to be twice than that in the rest of the Galaxy
\nocite{1994MNRAS.270..861G}({Gray} 1994a). Confirming the
SNRs from the MGCS is important as it suggests a possible
correlation between the dense environment in the GC
region and higher SNR density.

Shell type SNRs can be distinguished from other diffuse
sources in our galaxy by their ring like morphology and
non-thermal synchrotron emission which is characterised by a
steep power law and polarised emission. The other class of
SNRs, known as the plerions, are characterised by their
filled centre or blob-like morphology, a comparatively flat
spectrum and at high radio frequencies, can show a much
higher linear polarisation fraction than the shell type
SNRs. However, due to changing orientation of the magnetic
field and large Faraday depolarisations within the sources,
polarised emission is not detectable for many SNRs and
therefore measurements of their low frequency flux and
spectral indices are often the only way to detect
synchrotron emission. The Giant Metrewave Radio-Telescope
(GMRT) \nocite{1991CuSc...60...95S}({Swarup} {et~al.} 1991)
with it's large collecting area and high resolution at low
radio-frequencies is an ideal instrument for such studies.
Systematic GMRT studies of SNRs in the Galactic Centre
region are in progress (e.g.,
\nocite{2000MNRAS.317..453B}{Bhatnagar} 2000);
and in this paper, we discuss the observations of four
suspected SNRs G3.1$-$0.6, G356.3$-$0.3, G356.6+0.1 and
G357.1$-$0.2 which were selected from the MGCS.

In $\S$2 of this paper, we discuss observational
strategy and calibration. The results of our
observations  have been described in $\S$3, and in
$\S$4, we discuss about the morphology, spectral index,
distance and other physical properties of the candidate
SNRs observed by us. In $\S$5, we present a list of
compact sources seen within the field-of-view and also
an estimate of their flux at various radio frequencies.
We draw the conclusions in $\S$6.

\section{Observations and data reduction :}
The GMRT consists of thirty antennas, each of diameter
45 metre, distributed over a region of about 25
kilometres, with fourteen of the antennas placed within
a diameter of about one kilometre and the remaining
arranged in 3 arms each of length 14 km, shaped as an
irregular Y. This arrangement provides the necessary
{\it uv} coverage for mapping both compact and extended
sources.  The ratio of the longest to the shortest
baseline is around 500 with the shortest projected
baseline being about 50 metres.

The present observations were made when the GMRT was
still being commissioned and often all the 30 antennas
were not available. Table 1 give details of our
observations.
The centre of Field II was set at  l=356.8$^{\circ}$,
b=$-$0.1$^{\circ}$ so that  all the 3 objects,
G356.3$-$0.3, G356.6+0.1 and G357.1$-$0.2 were observed
within the primary beam of the antennas (FWHM=88$'$ at
330 MHz). We have also carried out an HI absorption
study towards G357.1$-$0.2. All the observations were
carried out in the spectral line mode with 128
frequency channels, which is the default for the GMRT
correlator. For the 330 MHz observations, a bandwidth
of 16 MHz was used while for the 21 cm spectral line
observations, which needed better frequency resolution,
the total bandwidth was reduced to 4 and 2 MHz, giving
a frequency resolution of 32 and 16 kHz respectively.
For the 330 MHz observations, 1830$-$36 was observed
initially for 20 minutes as a flux and bandpass
calibrator. Long term monitoring of this calibrator at
843 MHz using MOST has shown the modulation index
(ratio of the standard deviation to the mean value of
the flux density from the different epochs) to be 0.016
\nocite{2000PASA...17...72G}({Gaensler} \& {Hunstead} 2000)
which is well within the expected uncertainty of $\pm$ 15\%
in determining our absolute flux scale 
\nocite{1977A&A....61...99B}({Baars} {et~al.} 1977).
From the VLA calibrator manual (Perley 2001) we have taken
the flux density of this calibrator as 28 Jy at 330 MHz and
6.9 Jy at 1420 MHz.
The amplitude and phase calibrators were observed every
30 minutes. For Field I, the source 1830$-$21 was used
as both phase and amplitude calibrator.  For Field II,
the source 1822$-$096 was observed as the amplitude
calibrator and 1701$-$299 which was nearby at the same 
declination as the field being observed, was used as
phase calibrator.  Since, at the time of these
observations, automatic measurements of system
temperature ($T_{sys}$) were not implemented, the
$T_{sys}$ towards the calibrators and the SNR
fields were estimated using the 408 MHz all sky map
\nocite{1982A&AS...47....1H}({Haslam} {et~al.} 1982)
and then scaled to 330 MHz assuming a spectral index of
$-$2.6. The measured visibilities were corrected for this
$T_{sys}$ variations towards various sources (the typical
correction factors between the target source and the
secondary calibrator was $\sim$3). In this paper, we follow
the convention of F($\nu) \propto \nu^{\alpha}$, i.e., for steep
spectral index, the magnitude of $\alpha$ will be negative.
\begin{table*} 
\begin{minipage}{125mm}
\caption{Details of our observation} 

\baselineskip 20 pt

\begin{tabular}{|c c c c c c c|}
\hline
Galactic coord. &  RA & Dec & Frequency  & Date & Observing & Available \\
of the pointing &(J2000) & (J2000) & (MHz) &    & time (hours) &  antennas \\
Centre \it{(l, b)} &     &         &       & &         &           \\
\hline

003.2$-$1.0 & 17 57 00.1 & $-$26 40 01 & 330 & 1999 Jun 25 & 6 &   27 \\
356.8$-$0.1 & 17 38 11.4  & $-$31 42 26 & 330 & 1999 Aug 01 & 3.5 & 24 \\
357.2$-$0.2 & 17 39 40.6  & $-$31 28 01 & 1420.6 & 2000 Feb 26 & 3.5 & 15 \\
356.9+0.1 & 17 37 44.2  & $-$31 31 15 & 1420.3 & 2000 Oct 07 & 0.5 & 26 \\
\hline
\end{tabular}
\end{minipage}
\end{table*}
\subsection{Calibration}

The data were processed within AIPS using standard
programs. Bad data (interference in some frequency
channels, spikes and drop outs due to the electronics,
etc)  were identified and flagged using FLGIT.  The
data were  also checked for bad antennas or baseline
based problems which were also flagged. The spectral
visibility data were bandpass calibrated. For the
330 MHz data, a pseudo-continuum database of 4
frequency channels (each of width 2 MHz) were made from the
central 64 frequency channels of the observed 16 MHz band.
This was adequate to avoid bandwidth smearing within the
primary beam. Images of the fields were formed by Fourier
inversion and Cleaning (IMAGR).  The initial images were
improved by self-calibration. CLEAN maps of the  
compact sources in the field made by using only the
data with {\it uv}-distance greater than 150 $\lambda$ were
used to self-calibrate the data using CALIB with a {\it
uv}-range of 0.8$-$15 k$\lambda$.

For a better imaging of the extended structure in
the field G3.2$-$1.0, we made a high resolution map
with a lower cutoff of the {\it uv} range at 400
wavelengths.  The Fourier transform of the CLEAN
components of the sources in this map was then
subtracted from the {\it uv} data and one more map
was made with no lower cutoff in the {\it uv} range.
This map was then deconvolved using MEM. The
CLEAN components which were subtracted earlier were added back
to the resulting image to produce the final image of
the SNR. This technique has typically yielded an
RMS noise one and half times smaller than that
obtained using the standard CLEAN.  To minimise the
effect of non-coplanar baselines, the high resolution CLEAN
images were made for both the fields at 330 MHz using
the multi-facet technique available in the newer
versions of the AIPS task 'IMAGR'. 
However, to make the images of the field-II, the above
procedure was not required and we used standard CLEAN
to remove the side-lobes of the synthesised beam. 
All the GMRT maps presented in this paper have been
corrected for the primary beam pattern of the antennas.

We have also performed an HI absorption study with the
GMRT towards the candidate SNR G357.1$-$0.2 and the
compact source G356.9+0.1. 1748$-$253 was  used as
the phase calibrator and 3C287 as the bandpass
calibrator. The bandpass pattern of the
antennas change appreciably as a function of frequency.
Therefore, we have not used frequency
switching for the bandpass calibration, but, have
chosen a bandpass calibrator (3C287) with a high
Galactic latitude, so that the effect of Galactic HI
absorption on it's spectra is less than 1\%
\nocite{1978ApJS...36...77D}({Dickey}, {Terzian} \&
{Salpeter} 1978). We observed 1830$-$36 as the flux
calibrator.
The HI absorption profiles towards G357.1$-$0.2 and
G356.9+0.1 were obtained using 128 frequency channels
over a bandwidth of 4 and 2 MHz respectively. The
resultant velocity resolution is 6.6 km.s$^{-1}$
towards G357.1$-$0.2 and 3.3 km.s$^{-1}$ towards
G356.9+0.1. To make the channel maps, standard AIPS
task (UVLSF) was used to subtract a DC term
corresponding to the continuum from the visibility
data. The GMRT has an FX correlator, for which 'Gibbs
ringing' due to any sharp feature in the spectrum dies
away much faster (sinc$^2$ response to a sharp pattern,
which has a peak side-lobe of 5\%) as compared to an XF
correlator.  Therefore, we have not applied any
spectral smoothing to our data.  Also, the variation of
the line frequency introduced by the earth's rotation
during the observing period being much smaller than the
frequency channel width, we have not applied any
Doppler corrections.

\section{Results}

\begin{center}
\begin{figure}
\hbox{
\psfig{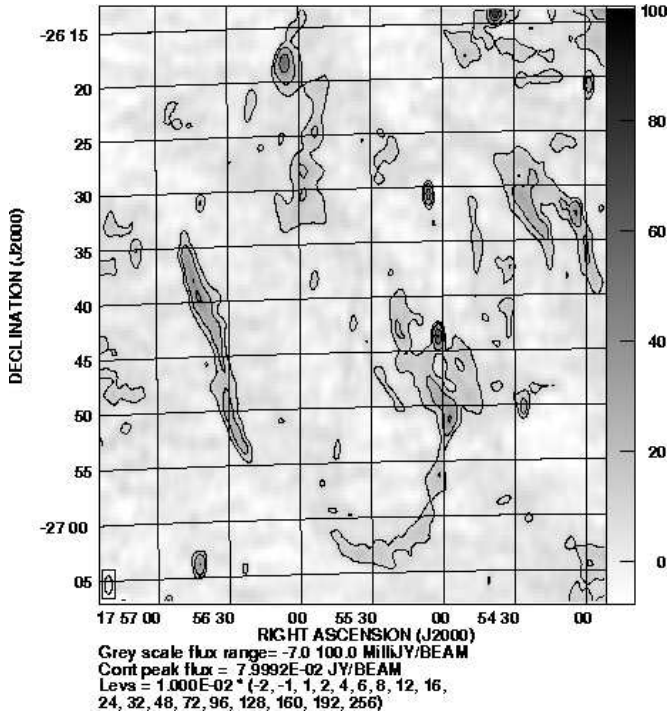}
}
\caption{MOST image of G3.1$-$0.6 at the original
resolution of 1.6$' \times$ 0.7$'$ along PA 0.1$^{\circ}$. The RMS noise is 4.5
mJy/beam.}
\label{g3.1.most}
\end{figure}
\end{center}
\vspace{-0.6cm}

\begin{center}
\begin{figure}
\hbox{
\psfig{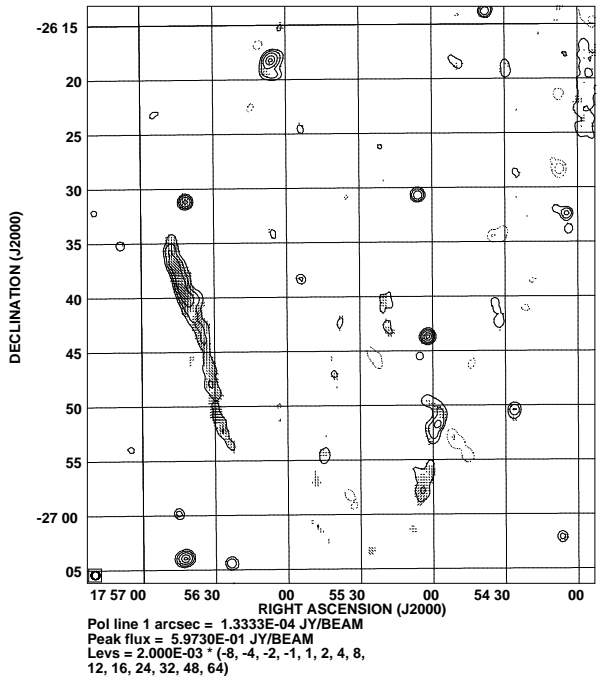}
}
\label{g3.1.nvss}
\caption{NVSS map of G3.1$-$0.6 at the original resolution of 45$^{''}$. 
The filamentary structure at the eastern side is
polarised. The RMS noise in the map is 0.5 mJy/beam.}
\end{figure}

\begin{figure}
\label{g3.1.hr}
\hbox{
\psfig{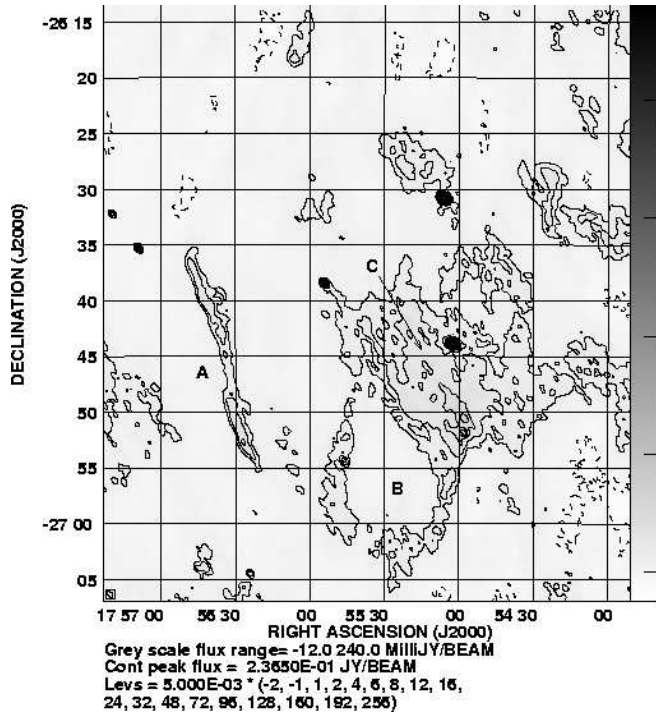}
}
\caption{GMRT high resolution image of G3.1$-$0.6 at 330 MHz. The resolution is 
37$^{''} \times $25$^{''}$ along PA 41$^{\circ}$ and the RMS noise in the map is 1.7 mJy/beam.}
\end{figure}
\end{center}
\begin{center}
\begin{figure}
\hbox{
\psfig{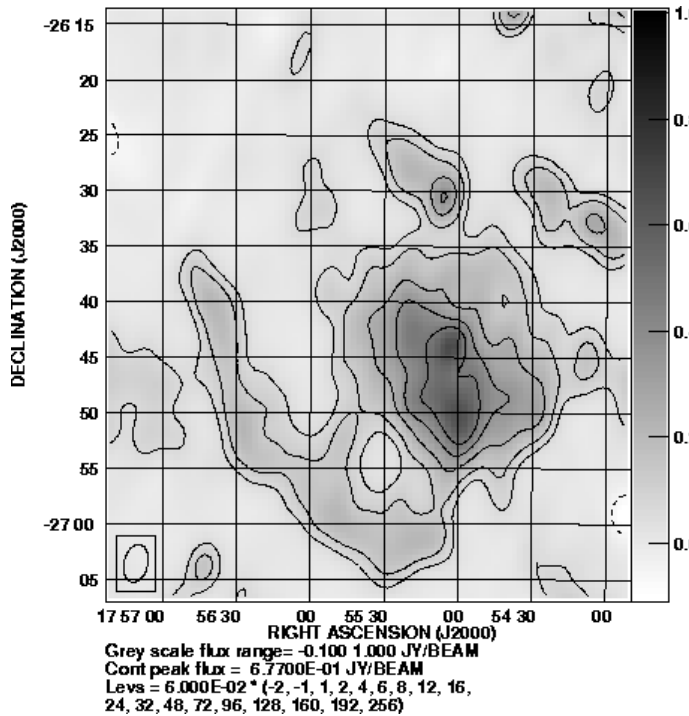}
}
\label{g3.1.lr}
\caption{GMRT low resolution image of G3.1$-$0.6 at 330 MHz. The resolution is 
3.4$' \times$ 2.1$'$ along PA $-$13$^{\circ}$ and the RMS noise in the map is 12 mJy/beam.}
\end{figure}
\end{center}
\vspace{-0.75cm}

\subsection{Field I:} 
The candidate SNR G3.1$-$0.6 identified during the MOST 843
MHz survey is located in this field.  In the MOST (Fig. 1)
and NVSS maps (Fig. 2), which are at higher frequencies, the
source appears as a collection of filamentary structures,
but, in the GMRT maps at 330 MHz (Figs. 3 and 4) extended
emission that connects up the filaments which form part of a
shell is clearly seen even in the high resolution maps. The
overall shape of the SNR is best seen in the GMRT low
resolution map in Fig. 4 which shows that the radio shell is
visible all around except in the northeast of the object.
There is general agreement between the structures seen in
the MOST and GMRT maps. However, the 330 MHz map also shows
appreciable diffuse emission towards the west (feature
marked C in Fig. 3). The eastern part of the shell as
evident in the low resolution image (Fig. 4), seems to be
quite narrow and appears like a filament in our high
resolution image (feature marked A in Fig. 3). This
filamentary structure is also seen in the NVSS
\nocite{1998AJ....115.1693C}({Condon} {et~al.} 1998) map
(Fig. 2). However, the NVSS maps are produced from snapshot
observations and due to a comparatively poor {\it
uv}-coverage in their data, the extended structures were not
properly imaged in their map.  The GMRT 330 MHz maps
indicate an additional small shell like feature at the
southern part of the larger shell that can be clearly seen
in Fig. 3 (feature marked B). 

From the GMRT image, the total flux density of this object
at 330 MHz is 14 $\pm$ 3 Jy, while at 843 MHz, the flux
density is reported to be 6.5 Jy
\nocite{1994MNRAS.270..847G}({Gray} 1994b). However, flux
estimated from both maps can suffer from zero-spacing
problems due to the large size of (28$'\times 49'$) of
G3.1$-$0.6. However, the filamentary structure near the
eastern side is sufficiently small (Fig. 3), so as not to
suffer from any significant zero-spacing problem.  The
spectral index estimated for the filamentary structure is
$-$0.62$\pm$0.12, which is relatively steep for an SNR.
This structure shows about 20\% polarisation in the NVSS
map. The diffuse emission at the western part of the remnant
also has a steep spectral index. We have examined the 8.35
GHz NRAO single dish map (Langston et al., 2000) of this
region.  Since we could not identify it, we can place an
upper limit of 2 Jy for this object. Hence, the spectral
index of the object is steeper than $-$0.64. We summarise
the observed parameters of G3.1$-$0.6 in Table 2.
It should be noted that the co-ordinate of this object in
Table 2 is different from \nocite{1994MNRAS.270..847G}{Gray}
(1994b). GMRT maps indicate that the centre of this object
is located further south of the co-ordinate given for this
object in \nocite{1994MNRAS.270..847G}{Gray} (1994b).
From Table 2, its estimated Galactic Co-ordinate is,
l=3.1$^{\circ}$, b=$-0.7 ^{\circ}$. However, as the object
has already been designated as G3.1$-$0.6, following the IAU
convention, we do not intend to rename it and will continue
to use it's name as G3.1$-$0.6 in the rest of this paper.
\vspace{0.6cm}

\begin{table*}
\begin{minipage}{130mm}
\caption{Observed parameters of G3.1$-$0.6 and G356.8$-$0.0}
\baselineskip 20 pt
\begin{tabular}{|c c c c c c c c|}
\hline
Object &  RA & Dec & Size & Type & S$_{843 MHz}$ & S$_{330 MHz}$ &
Spectral index \\
name (G\it{l$\pm$b}) & (J2000) & (J2000) & ($'$) &  & (Jy) & (Jy) & \\
\hline
G3.1$-$0.6 & 17 55 40 & $-$26 39 30 & 28$\times$49 & Shell & 6.5 & 14 $\pm 3$  & $\le$
$-$0.64 \\

G356.8$-$0.0 & 17 38 00 & $-$31 37 41 & 52$\times$72(?) & Shell & -- & 30 & $\le$ $-$0.34 \\
\hline

\end{tabular}
\end{minipage}
\end{table*}
\subsection{Field II:}

This region has been investigated previously by 
\nocite{1994MNRAS.270..847G}{Gray} (1994b) using the MOST at
843 MHz.  Three candidate SNRs, G356.3$-$0.3, G356.6+0.1 and
G357.1$-$0.2 were reported and the MOST image of these
objects are shown in Fig. 5.  This field is much more
complex than Field I and further there are a number of
complications due to the proximity of the 'Tornado nebula'
which has a large flux density.  The GMRT low resolution map
of this field at 330 MHz is shown in Fig. 6. The prominent
features of this field are (a) the strong source at the
northeastern edge of the map which is identified with the
Tornado Nebula, (b) a strong unresolved source G356.9+0.1
(RA=17h37m44.0s, Dec=$-$31$^{\circ}$31$'$14${''}$ (J2000)) which
with higher resolution has three components (Fig. 10) and shows a
non-thermal spectrum (Table 5), indicative of a
background extragalactic source and (c) two extended objects
which have considerable fine structure. The large extended
emission is located to the south of the extragalactic source
G356.9+0.1, which appears to be curved has been detected for
the first time. 
The other extended emission to the east of the
extragalactic source coincides with the MOST candidate SNR
G357.1$-$0.2, though the extent of its emission at 330 MHz
is larger than that in the MOST maps 
This second extended source almost touches the southern
emission, and they appear to be part of a single large
shell, the northern part of which is missing. 

It should be noted that interferometric observations which
lack zero and short-spacings, can preferentially pick-up
small scale structures from extended Galactic emission and
hence can show structures in the map which are not genuine
entities.
These small scale variations in the Galactic background
emission generally create structures with both positive and
negative flux-densities in the map plane, which finally
increases the RMS noise in the map. However, in Fig. 6, the
first contour level is 0.1 Jy/beam, 5 times higher than the
RMS noise and at that level, no extended region of negative
flux-density can be seen. Therefore, we believe that the two
extended objects seen in our map are genuine in nature.

In the absence of the northern part of the shell, we can
only speculate about it's actual extent to the north. If the
conceived shell is formed by the line joining G357.1$-$0.2
with G356.6+0.1, then, the shell has a size of $52' \times
55'$. However, it appears to us that the northern extent of
the shell is upto the object seen at RA=17h38m00s,
Dec=$-$31$^{\circ}$05$'$, and consequently the shell has a size of
$52' \times 72'$ centred at G356.8$-$0.0. We designate this
conceived shell by its galactic co-ordinate of G356.8$-$0.0.
The flux density calculated over the two extended objects at
330 MHz is 30Jy.  However, this should be taken with caution
since we are likely to underestimate its flux due to missing
short spacings.

\begin{center}
\begin{figure}
\hbox{
\psfig{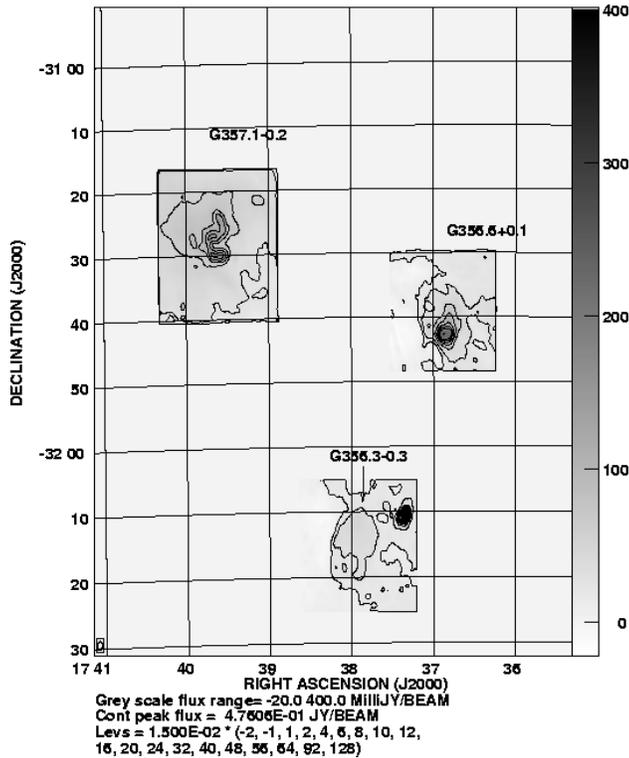}
}
\label{most.g356}
\caption{A composite of the MOST 843 MHz maps of the 3
suspected SNRs G356.3$-$0.3, G356.6+0.1 and G357.1$-$0.2 at
the original resolution of 1.3$' \times 0.7'$ along PA
0.1$^{\circ}$. The RMS noise in the map is $\approx$ 6
mJy/beam.} 
\end{figure}
\end{center}
\vspace{-0.6cm}
\begin{center}
\begin{figure}
\hbox{
\psfig{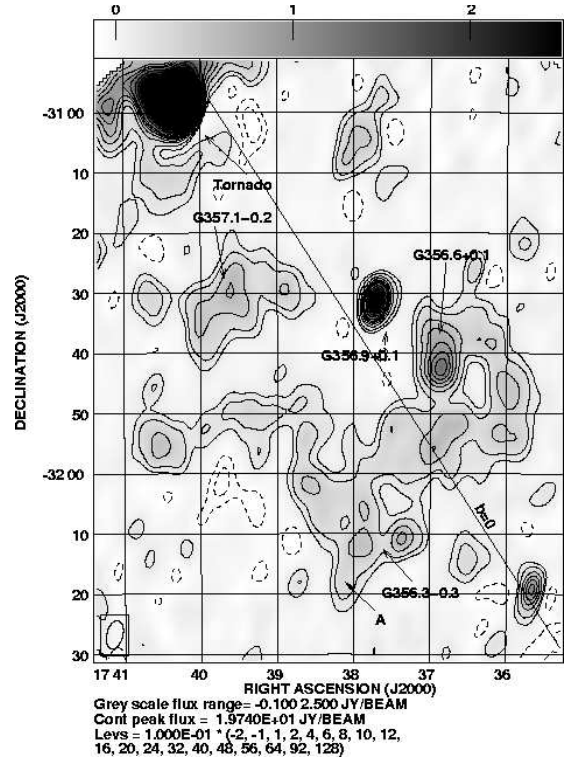}
}
\label{g356.lr}
\caption{GMRT low resolution map of the Field II at 330 MHz. 
The resolution is 4.8$' \times$ 2.7$'$ along PA $-$14$^{\circ}$ and the RMS noise in
the map is 20 mJy/beam.}
\end{figure}
\end{center}

A composite of the MOST maps within this field is given in Fig.
5, along with the NVSS map of this region (Fig. 7). The
MOST suffer from zero-spacing problems for objects
larger than half a degree which, coupled with further
dynamic range limitations due to the nearby strong
source (Tornado nebula) could have resulted in their
inability to image the large extended structure
(G356.8$-$0.0) and to identify smaller scale structures
within the extended emission as separate SNR candidates
(equivalent to a 'high frequency spatial filtering').
The NVSS map of the field has serious short spacing problems
and it is not surprising that the extended emission is not
seen in it. Single dish maps at high frequencies do not have
this problem, but since the extended emission has low surface
brightness, it is often difficult to detect it. The extended
emission is not seen in the 8.35 and 14.35 GHz NRAO single
dish maps (Langston et al. 2000), from which we set an upper
limit on the flux density of 9.4 Jy at 14.35 GHz and 10Jy at
8.35 GHz. It should be noted that Langston et al. used a
median filter of size 67$'$ to remove the extended
background emission. 
However, since the area enclosed by the shell is
significantly smaller than the filter diameter, removing the
extended background is not expected to cause any appreciable
underestimation of the flux density. If we combine the 330
MHz results with the upper limits from Langston et al., we
find that the spectral index of the shell like structure is
steeper than $-$0.34. Based on the above, we suggest that
G356.8$-$0.0 could be an old SNR shell, which has started
breaking up near G357.1$-$0.2, possibly due to interaction
with the surrounding ISM. 

The observed parameters of this shell have been summarised
in Table 2. We compare the observed properties of these
sources at different frequencies in the discussion below.

\begin{center}
\begin{figure}
\hbox{
\psfig{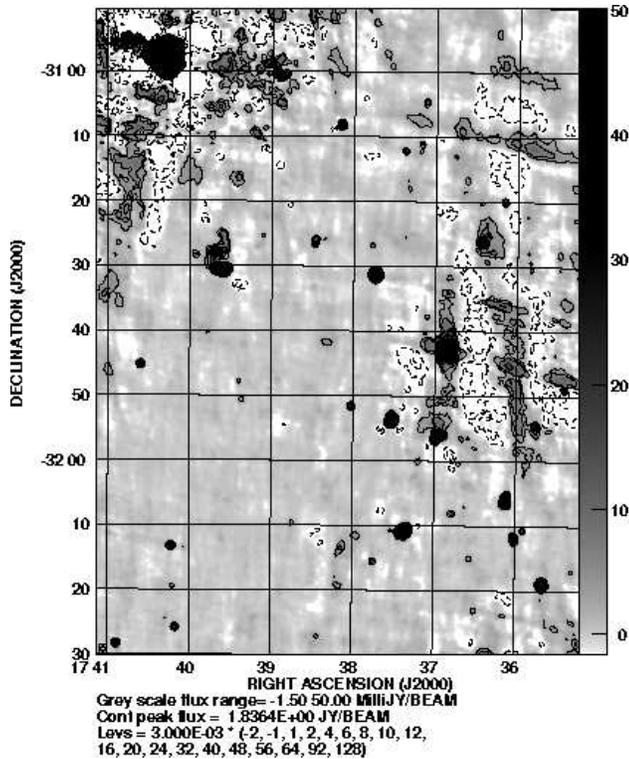}
}
\label{nvss.g356}
\caption{NVSS 1.4 GHz map of the Field II at the 
original resolution of 45$^{''}$. The RMS noise in the
map is 0.8 mJy/beam.}
\end{figure}

\begin{figure}
\hbox{
\psfig{file=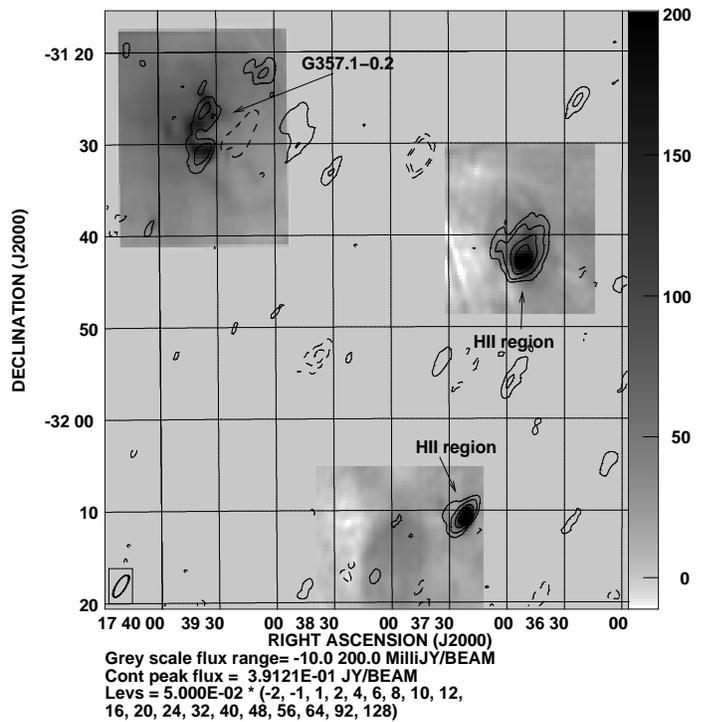,clip=,height=10cm}
}
\label{g356.comp}
\caption{GMRT 330 MHz map of the field-II in contour
overlaid on the MOST 843 MHz Gray scale map of the
same region. The MOST Gray scale map has been presented at the
original resolution of 1.3$' \times 0.7'$ along PA
0.1$^{\circ}$. The GMRT 330 MHz contour map has a
resolution of 2.8$' \times$ 1.0$'$ along PA
-33$^{\circ}$ and the RMS noise in the map is 17
mJy/beam. The GMRT map was made by first
subtracting the CLEAN components of the compact sources
from the {\it uv} data and a lower {\it uv} cutoff of
230 $\lambda$ was applied to the data before
Fourier Transforming and CLEANING.} 
\end{figure} 
\end{center}
\vspace{-0.6cm}

\noindent (i) G356.3$-$0.3: \\ In the MOST image, this SNR
candidate appears as a circular object of size 10$'$.
However, the GMRT map as shown in Fig. 8,
there is no discrete source at this position. In the GMRT
low resolution map (Fig.  6), this object is located on the
extended structure, which we have interpreted as part of a
larger shell. 
In the NVSS map, a small partial shell structure can be
identified (not clear in Fig. 7) near the eastern boundary
of G356.3$-$0.3 (RA=17h38m and Dec=-32$^{\circ}$15$'$), which
also coincides with the southern edge of
G356.8$-$0.0 (feature marked 'A' in Fig. 6) in the 330 MHz
map.  We suggest that the partial shell structure in the
NVSS map is due to a lack of short {\it uv}-spacings, which
results in a 'high frequency spatial filtering' of the
feature 'A' shown in Fig. 6.
Based on the above, we conclude that G356.3$-$0.3 is
likely to be part of the larger shell (G356.8$-$0.0) like
structure. The total flux density calculated for this object
from the MOST image after background subtraction is 2 Jy.
From the same region, the estimated flux density at 330 MHz (Fig.
6) is $\approx$1.5 Jy, and is 0.1 $\pm$ 0.02 Jy at 1.4 GHz
(Fig. 7).

\noindent (ii) G356.6+0.1 :\\
This suspected SNR candidate
\nocite{1994MNRAS.270..847G}({Gray} 1994b) appears as a
featureless elliptical emission with a central compact HII
region in the MOST image. As the extended emission
surrounding the HII region is not visible in the IRAS
images, this was classified as an SNR candidate. The GMRT
image also is similar but the extended emission seems
associated with the large shell like structure discussed
before.  After excluding the HII region, we estimate a flux
density of 2.2 Jy for this source at 330 MHz. From the MOST
image, the corresponding flux density is 1.7 Jy. In NVSS,
the central HII region is visible, but the extended emission
has stripes due to the lack of shorter {\it uv} spacing
required to image it.  The flux density of the central HII
region as estimated from the NVSS, the MOST and the GMRT
image do agree to within $\pm$ 20\%, the estimated flux
density being 1.2$\pm$0.2 Jy at 330 MHz. As the extended
emission around the HII region has a non-thermal spectrum
and appears to be part of the large shell like structure
G356.8$-$0.0, we suggest it to be a smaller-scale structure
within G356.8$-$0.0 and not a discrete SNR. The central HII
region can well be an unrelated background object.

\noindent (iii) G357.1$-$0.2:\\ 
In the 843 MHz map, G357.1$-$0.2 shows a 
bi-annular structure of a few arc-minutes with a halo of
emission surrounding it. In the 330 MHz low resolution
map shown in Fig. 6, the bi-annular structure is not
distinguishable, but there is an extended emission of
size ~15$'$ at the same place. To search for this
structure, we first subtracted the strong sources in
the region from the {\it uv} data and resolved out the
extended emission by using a lower {\it uv} cutoff of 230
$\lambda$, while re-imaging the region. The resultant
contour map is shown in Fig. 8, where, the MOST map has
been overlaid on it in gray scale.  There is indeed
seen some emission at the position of the bi-annular
structure of the MOST image (The nearby negative is the
result of missing zero-spacing flux density resulted
from our attempt of resolving out the extended emission
of size ~15$'$). From the GMRT image, we estimate a
flux density of 0.49 $\pm$ 0.15 Jy for this object.
From the MOST image, we estimate a flux density of 1.2
Jy at 843 MHz for this object after background
subtraction. 

As noted by \nocite{1994MNRAS.270..847G}{Gray} (1994b), this
region of the MOST image is suspected to be affected by the
base-level fluctuations caused by the nearby strong source
(Tornado nebula).  As a check, we have estimated the flux
density of the surrounding 15$'$ halo region from their 843
MHz map to be about 10 Jy.  But, from the GMRT image, the
estimated flux density is only 5.0$\pm$1.0 Jy at 330 MHz
(Fig. 6). Further, the upper limit of the flux density
measured from the 8.35 GHz NRAO single dish map (Langston et
al. 2000) is 1.0 Jy for this region. The above measurements
clearly indicate that there is a problem in explaining the
flux density as measured at 843 MHz, as, neither an HII
region (any significant free-free absorption at 330 MHz can
be ruled out) nor a mixture of thermal and non-thermal
emission can explain the estimated flux density at 843 MHz.
Therefore, it seems that the flux density measurements of
this region from the MOST image is badly affected by the
background uncertainty and hence we have not used the flux
density of 1.2 Jy calculated for the bi-annular structure at
843 MHz any further in this paper.

From our HI data of this source we have made a continuum map
using the line free channels (Fig. 9) which gives us a
flux density of 0.67$\pm$0.1 Jy for this object at 1.4
GHz. From the NVSS map at 1.4 GHz
\nocite{1998AJ....115.1693C}({Condon} {et~al.} 1998), the
estimated flux density is 0.45$\pm$0.03 Jy. To resolve this
difference, we re-mapped the same region by using the
calibrated {\it uv} data of the NVSS (available in the web).
The estimated flux density for this object after applying
the primary beam correction is found to be 0.7 $\pm$
0.1 Jy, in agreement with our measurement.

This source has also been observed by
\nocite{1996ASP...102.443}{Gray} (1996)
using the VLA at 5 GHz (DnC array) and 1.4 GHz (CnB
array).
This object shows several filamentary structures.  Between 5
and 1.4 GHz, most parts of the structure show a spectral
index of $-$0.5, though, there is a variation of spectral
index from $-$0.3 to $-$0.6. The structures seen at 5 GHz
are highly polarised and display a polarisation fraction of
upto 60\% \nocite{1996ASP...102.443}({Gray} 1996).
Therefore, the emission mechanism from the source is
synchrotron in origin. We further discuss it's properties in
the following sections.

\begin{center} \begin{figure}
\hbox{
\psfig{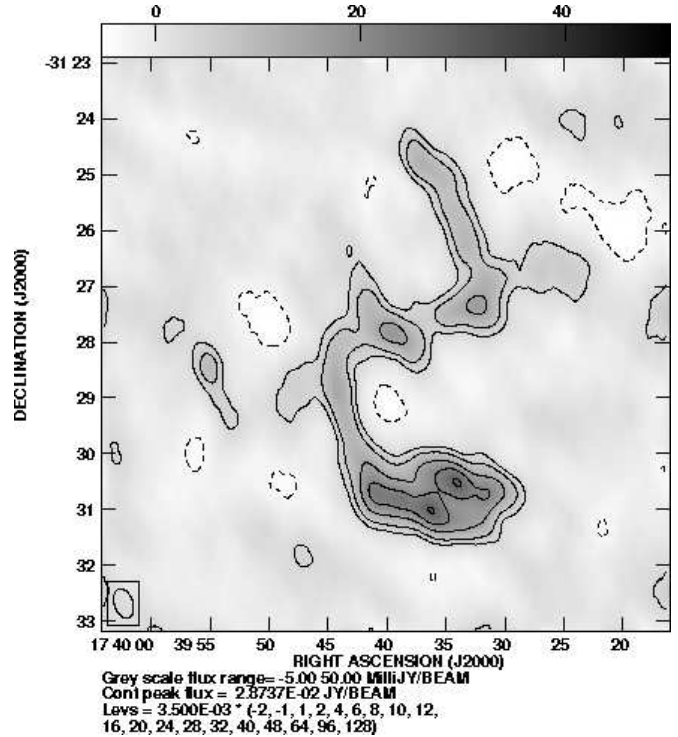}
}
\caption{GMRT 21 cm map of G357.1$-$0.2. The resolution
is 33$^{''} \times$ 20$^{''}$ along PA 19$^{\circ}$ and the RMS noise is 1.5 mJy/beam.}
\label{g357.vla}
\end{figure}
\end{center}
\vspace{-0.6cm}
\begin{center} 
\begin{figure}
\hbox{
\psfig{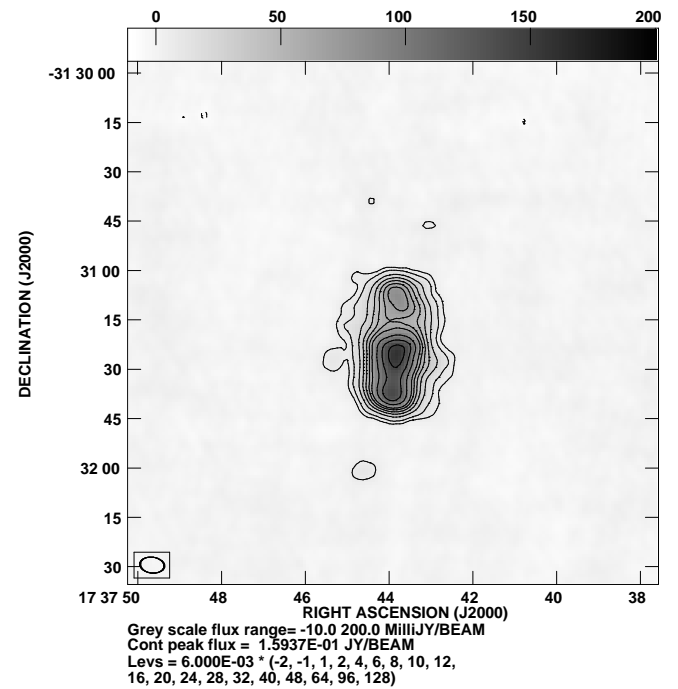}
}
\label{g356.9.cont}
\caption{GMRT continuum map of G356.9+0.1 at 21 cm. The
resolution is 7.4$^{''} \times$ 4.8$^{''}$ along PA 83$^{\circ}$ and the RMS
noise is 1.7 mJy/beam.}
\end{figure}
\end{center}
\vspace{-0.6cm}

\begin{center} 
\begin{figure}
\vbox{
\hbox{
\psfig{file=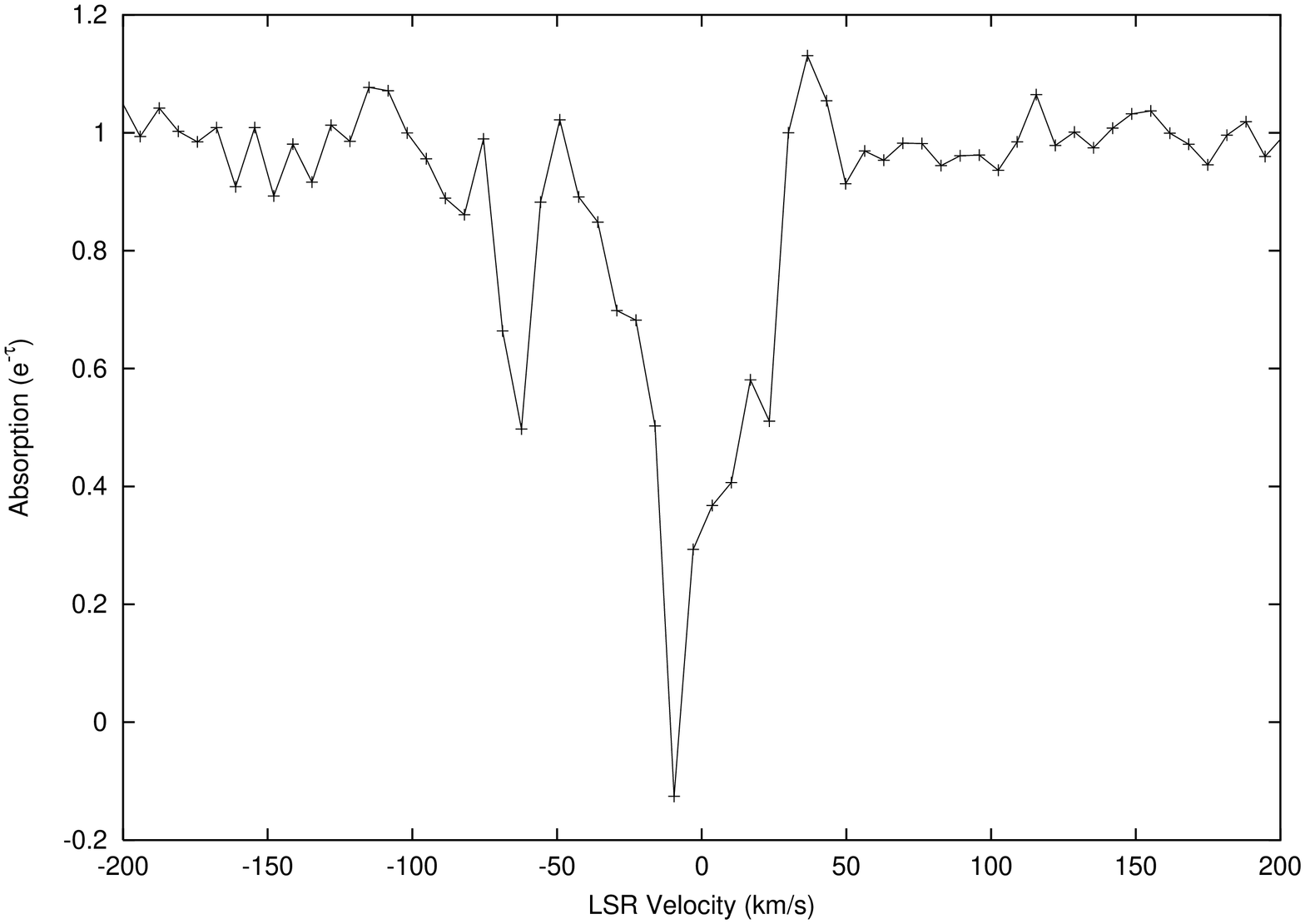,angle=270,clip=,height=5.7cm}
}
\caption{HI absorption spectra towards G357.1$-$0.2.
The spectra was made by integrating over a rectangular
region of the line maps, where strongest continuum
emission from the source is observed.}
\label{g357.1.hi.abs}
\vspace{+0.0cm}

\hbox{
\psfig{file=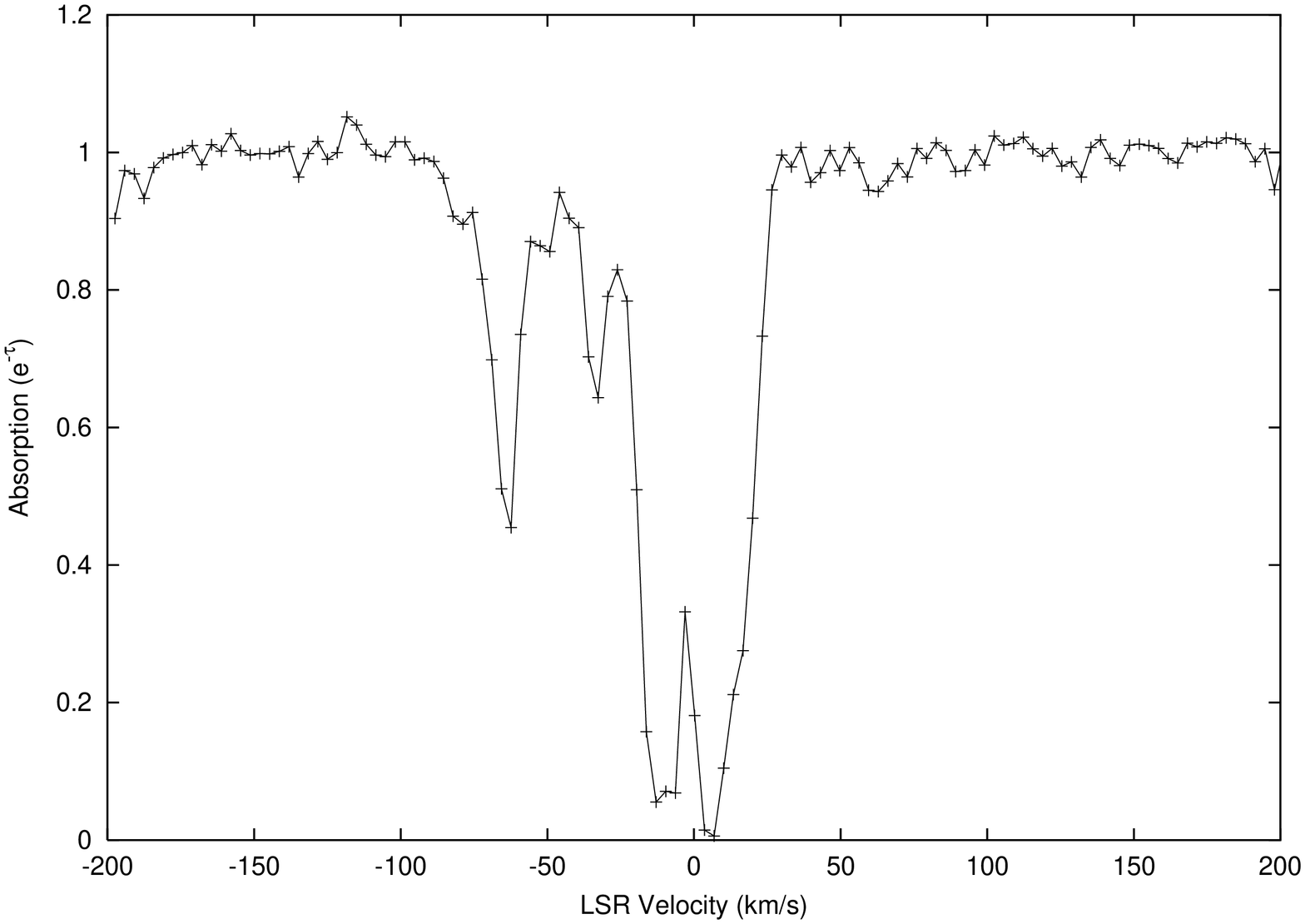,angle=270,clip=,height=5.7cm}
}
}
\caption{HI absorption towards G356.9+0.1}
\label{g356.9.hi.abs}
\end{figure}
\end{center}

\subsection{HI absorption towards G357.1$-$0.2 :}

In Fig. 11 is shown the normalised spectra of the
source, where, the y-axis represents the estimated
absorption ($e^{-\tau}$, where, $\tau$ is the optical
depth).
Within $\pm$20 km.s$^{-1}$ of the LSR velocity, the
RMS noise is 0.07 of the absorption profile (i.e.,
$e^{-\tau}$), but, for the frequency channels located
outside the above LSR velocity range, the RMS noise
is 0.035. The HI emission from the Galactic plane
increases the system temperature for the frequency
channels located within $\pm$20 km.s$^{-1}$ of the LSR
velocity, causing an increase in the RMS noise for
the frequency channels located in the above range. In
our map of G357.1$-$0.2, HI emission was also found to
increase the confusion limit for the frequency channels
located within the above velocity range. To reduce the
effect of HI emission from the extended structure along
the line-of-sight gas, we applied a lower {\it uv}
cutoff of 500 $\lambda$ before imaging the
spectral-line {\it uv} data. We have also tried to
identify any small scale HI emission associated with
this source from our line maps.  However, no evidence
was found in this regard. The source G357.1$-$0.2 is
seen in absorption near the LSR velocity $-$10 and $-$62
km.s$^{-1}$. The absorption near the $-$62
km.s$^{-1}$ arises from the '3 kpc arm' (for which
$V_{LSR} \approx -$60 km.s$^{-1}$) indicating that
this source is located at least at a distance of more
than 6 kpc from us (The GC is $\approx$ 8.5 kpc away
from us). The absorption near the $-$10 km.s$^{-1}$ is
quite narrow with the velocity width (FWHM) $\sim$ 28
km.s$^{-1}$ (Fig.  11). For comparison, we have also
observed the compact object G356.9+0.1 located near the
centre of our 330 MHz map, in HI absorption at 1420
MHz. This object appears to have three components in
our high resolution 1.4 GHz continuum map (Fig. 10) and
is likely to be an extragalactic source. The absorption
spectra towards this source is shown in Fig. 12 (the
RMS noise in the spectra is 0.02 for the frequency
channels located within $-20 \ge V_{LSR} \le 20$, and
is $\approx$0.01 outside this range.). As expected,
this object is seen in absorption against the '3 kpc
arm'. At LSR velocity of $-$33 km.s$^{-1}$, one more
absorption line can be seen against this source which
is not observed towards G357.1$-$0.2.  The HI
absorption profile shows a large velocity width (FWHM
$\sim$40km.s$^{-1}$) near the LSR velocity of 0
km.s$^{-1}$ towards this object. 

Similar broad absorption features have been observed
for the extragalactic sources seen through the Galactic
Centre region, e.g., 1748$-$253, G0.537+0.26
\nocite{1983ApJS...53..591D}({Dickey} {et~al.} 1983) and G359.87+0.18
\nocite{1999ApJ...515..196L}({Lazio} {et~al.} 1999) which are believed to be
caused by an anomalous velocity cloud located behind the
Galactic Centre. As the absorption FWHM towards
G357.1$-$0.2 is significantly lower as compared to
G356.9+0.1, we think that this object is likely to be a galactic
source. However, the full width of absorption towards
this object is quite similar to G356.9+0.1, which
suggests that G357.1$-$0.2 may be embedded within the
+ve velocity cloud located behind the Galactic Centre,
which gives rise to much deeper absorption profile towards
G356.9+0.1 because of the longer line of sight towards
this extragalactic source.

\section{Discussions:} 

\subsection{G3.1$-$0.6 (Field I):}
\subsubsection{Distance:}
From the present observations, only the $\Sigma$-D relation
can be used to estimate it's distance. We used the modified
$\Sigma$-D relation ($\Sigma_{{\rm1\ GHz}} = 2.07 \times
10^{-17} \times {\rm D}^{-2.38}$
W.m$^{-2}$.Hz$^{-1}$.sr$^{-1}$) by
\nocite{1998ApJ...504..761C}{Case} \& {Bhattacharya} (1998),
which is expected to yield distance to individual SNRs with an
error $\sim$40\%. Using a flux density of 6 Jy for this
object at 1 GHz and a diameter of 39$'$, we estimate a
distance of 7.7 kpc for this object, which indicates that it
can even be located near the Galactic Centre region.

It should be noted that the $\Sigma$-D relation is essentially
derived by assuming the environments of the SNRs into which
expansion occurs are similar. It is believed that the
$\Sigma$-D relation overestimates the diameter of SNRs
located at higher Galactic latitudes
\nocite{1979MNRAS.187..201C}({Caswell} \& {Lerche} 1979).
This phenomenon is thought to be related with intrinsic
decrease of the surface brightness of the SNRs due to a
lowering of magnetic field at higher Galactic latitudes.
However, the magnetic field in the GC region is expected to
be much higher than the Galactic disk, which can result in a
higher intrinsic surface brightness of the SNRs in the
region. 
Hence, the use of $\Sigma$-D relation to estimate the
distance or linear-diameter of the SNRs located in the GC
region can result in underestimating these parameters.
For example, the distance estimated by using the $\Sigma$-D
relation \nocite{1998ApJ...504..761C}({Case} \&
{Bhattacharya} 1998) for the SNR Sgr A East (G0.0+0.0) is 3.3 kpc.
However, it is thought to be actually located behind the GC
(i.e., at a distance of $\ge$ 8.5 kpc)
\nocite{1989ApJ...342..769P}({Pedlar} {et~al.} 1989).

\subsubsection{Morphology:}

G3.1$-$0.6 shows two ring like structure in our high
resolution map with the smaller ring being located
towards the south. Possible explanations of the
observed morphology are (a) the SNR itself is the
combination of two separate objects, (b) this
morphology could be a result of the expansion of the
supernova blast-wave into a cavity having a bi-annular
structure created by stellar winds of the progenitor
system, (c) the ISM in which the SNR expands has
structures which are responsible for such a morphology.

Given the high density of SNRs in the Galactic Centre
region, the suggestion that two SNRs lie close to each
other is not unreasonable. There are at present 15
known SNRs within the central 50 square degree area of
the Galactic Centre region and the probability of the
centre of one the SNRs to be within 0.5$^\circ$ of another
is $\sim$7.5\%, which is not negligible. This
hypothesis can be discarded if the two shells can be
shown to have different distances, ages or spectral
index.  The spectral indices estimated from the 330 MHz
and the 843 MHz map for the two shells are not very
different  and we cannot rule out the two SNR
hypothesis, though, we prefer the case of an
anisotropic ISM as described below.

It is now known that the slow wind from the red
super-giant star (RSG) possess axial symmetry
\nocite{1996ApJ...472..257B}({Blondin}, {Lundqvist} \&  {Chevalier} 1996). 
\nocite{1987A&A...171..205M}{Manchester} (1987) 
proposed that expansion of an SNR into such a medium
can produce a double ringed morphology.  
\nocite{1998MNRAS.296..813G}{Gaensler}, {Manchester} \&  {Green} (1998) 
have shown for the SNR G296.8$-$0.3, the asymmetric
cavity as created by the RSG wind of the progenitor
star has a typical size of $\sim$5 pc. However, if
G3.1$-$0.6 is placed at a distance of 7 kpc, it would
have a linear size of 80 pc. Therefore, RSG wind can be
ruled out as one of the possible reason for the double
ring structure.

Finally, we consider the case of an anisotropic ISM being
responsible for the shape of the SNR. It
has been argued that the SNRs G166.0+04.3 (VRO
42.05.01) \nocite{1987ApJ...315..580P}({Pineault}, {Landecker} \&  {Routledge} 1987), 
G350.0$-$02.0 \nocite{1998ApJ...493..781G}({Gaensler} 1998) 
and G296.8$-$0.3 \nocite{1998MNRAS.296..813G}({Gaensler} {et~al.} 1998) 
represents the case of a supernova shock re-energising
a tunnel and then propagating on to the other side to
form a second shell. We find that the morphology of the
SNR G3.1$-$0.6 can also be explained on the basis of
the above model.  In order to explain the model, we
depict the main two ring structures in a schematic
diagram as shown in Fig. 13.

\begin{center}
\begin{figure}
\hbox{
\psfig{file=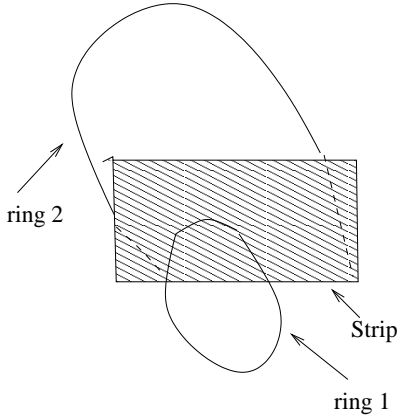,angle=0,clip=,height=5.5cm}
}
\caption{A sketch of the morphology of the SNR G3.1$-$0.6}
\label{g3.1.morphology}
\end{figure}
\end{center}
\vspace{-0.6cm}

(i) There is a smaller ring of emission ('ring 1'),
which in the absence of the other structures would
appear as a typical SNR (feature B
in Fig. 3.). 
(ii) A strip of emission running west from the eastern
side of the SNR (feature marked C and the weak emission
towards the eastern edge of Fig. 3) which,
we suggest corresponds to a tunnel of much
lower density. On encountering such a region, an
expanding shock will energise it, rapidly
propagating both across the tube and up and down 
it's length. As this process continues, electrons in the
walls are shock-accelerated and emit synchrotron
radiation \nocite{1987ApJ...315..580P}({Pineault} {et~al.} 1987) 
giving a linear structure as observed here.  This is
quite similar to the case of the SNR G296.8$-$0.3.
(iii) A comparatively larger ring ('ring 2'), located
north of ring 1 (corresponding to the large ring
structure in Fig. 4). Ring 2 is fainter than ring 1
(as seen in Fig. 3) and extends much beyond the
boundary of the ring 1 in the north. This ring  has a
poorly defined outer edge especially towards the north,
which suggests that it represents a region of break-out
into an adjacent cavity. This cavity itself could have
been created in the past by some other energetic
phenomena. In this case, the shock after being
re-energised in the low density tunnel, will expand
rapidly to take the shape of the cavity
\nocite{1986A&A...164..193B}({Braun} \& {Strom} 1986). 
For the SNRs G166.0+04.3 (VRO 42.05.01) and
G350.0$-$02.0, all the
above three components are located in the plane of the sky.
Whereas for G296.8$-$0.3, the explosion site, the low
density tunnel and the second shell all lie roughly
along the line of sight. We suggest that SNR G3.1$-$0.6
may be quite similar to the case of G296.8$-$0.3
\nocite{1998MNRAS.296..813G}({Gaensler} {et~al.} 1998), 
where all the three components
lie along the line of sight.

\subsubsection{Spectral index and polarisation:}
Non-thermal steep spectral index, polarised emission
and shell morphology clearly identifies G3.1$-$0.6
to be a shell type SNR. 
The detection of polarised emission suggests that the
magnetic field in the region is well organised. It
also indicates that the maximum rotation measure is less than 1000
rad m$^{-2}$ towards the SNR along the line of sight of medium.

\subsection{G356.8$-$0.0 (Field II):}
There are two extended curved emission seen in the map of
this region. Following the discussion in $\S$3.2, both
of these objects appears to be part of a single large shell
named as G356.8$-$0.0. Its non-thermal spectral index and a
partial shell like morphology suggests the large shell like
structure to be of an old SNR. However, due to its low
surface brightness, its flux density at higher frequencies
are poorly determined. As a result, its spectral index ($\le
-0.34$) is not well constrained.

With the available data, only the $\Sigma$-D relation
can be used to estimate it's distance.
With an assumed spectral index of $-$0.5, the estimated flux
density of this object is 20 Jy at 1 GHz. The modified
$\Sigma$-D relation as developed by
\nocite{1998ApJ...504..761C}{Case} \& {Bhattacharya} (1998),
yields an estimated distance of 4.4 kpc for this SNR
candidate. However, due to the short-spacing problem and the
inherent uncertainty in the $\Sigma$-D relation itself, this
distance may be uncertain by as much as a factor of three.
The large shell shows kinks at different places and also
seems to have broken up near G357.1$-$0.2, suggesting it to be
an old SNR. Assuming an expansion velocity of 50 km.s$^{-1}$
and it's angular size of 60$'$ at a distance of 4.4 kpc, we
use the \nocite{1974ApJ...188..501C}{Chevalier} (1974) 
formula t=1.5 x 10$^5$ D v$^{-1}$ to estimate it's probable age. The
calculated age is 2.4$\times 10^5$ years.
\begin{table*} 
\begin{minipage}{130mm}
\caption{Observed parameters of G357.1$-$0.2} 

\baselineskip 20 pt

\begin{tabular}{|c c c c c c c c|}
\hline
Galactic &  RA & Dec & Size & S$_{1.4 GHz}$ & S$_{330 MHz}$ &
Spectral & Spectral \\
Coords \it{($l \pm b$)} &(J2000) & (J2000) & ($'$) & (Jy) & (Jy) & Index & Index \\
   &          &              &           &        &       & (5/1.4) & (1.4/0.330) \\
\hline
357.1$-$0.2 & 17 39 38 & $-$31 28 57 & 8$\times$7 & 0.67$\pm 0.1$ &  0.49$\pm 0.1$  & $-$0.5 & 0.2
$\pm 0.2$ \\ 
\hline

\end{tabular}
\end{minipage}
\end{table*}
\subsection{Bi-annular source G357.1$-$0.2 (Field II):}
The flux density of this object at 330 MHz along with
it's flux density measured at 1.4 GHz shows a slightly inverted
spectral index of +0.2 $\pm$ 0.2. However, between 5
and 1.4 GHz, it has a spectral index of $\sim -$0.5.
This indicate a break in it's spectral index near 1 GHz
with a difference of 0.7 between it's high and low
frequency spectral index. The peculiar morphology makes
it difficult to draw unique identification for this
object. Highly polarised emission at 5 GHz, however,
shows the emission to be non-thermal in nature. As
noted by Gray (1996), this object can be identified
with (i) an extragalactic source, (ii) filamentary
structure as seen close to the GC or (iii) a Crab like
SNR. Compact cores of extragalactic sources
often show a flat spectral index, which turns
over at low radio-frequencies $\sim$ 100 MHz due to
the synchrotron self-absorption. However, the extended
components have a steep spectral index and shows
relatively low level of absorption only at very low
frequencies.  As this object is extended, our determination
of a turn over of it's spectra below 1 GHz  argues against
the object being a typical extragalactic source.  Also, it's
HI absorption spectrum suggests it to be galactic.
Filaments near the Galactic Centre have not been reported to
show any spectral break and therefore, it is unlikely to be
a Galactic Centre filament.  Within the observational error,
this object has a flat or slightly inverted spectrum at low
radio frequencies. A slightly inverted spectrum, if correct,
can be a result of line of sight free-free absorption by
diffuse foreground HII region.  We argue below that optical
depth ($\tau$) due to free-free absorption can't be more
than 0.2 at 330 MHz towards this object.

(i) Any line of sight structure of size larger than
40$'$ will get resolved out in our map. However, we can
estimate the free-free absorption coefficient by using
the 408 MHz map of \nocite{1982A&AS...47....1H}{Haslam} {et~al.} (1982), 
which gives a 
sky temperature of 440 K in this region at 408 MHz.
An extended HII region will
typically have an electron temperature (T$_e$) greater
than 4000 Kelvin, indicating $\tau \le$ 0.11 at 408
MHz, which is equivalent to $\tau \le$ 0.17 at 330 MHz.

(ii) HII region of size $\le$ 40$'$ and optical
depth $\ge$ 0.001 should be visible in our map above
the noise level. As nothing of that sort is visible, we
can rule out any perceptible line of sight absorption
by a moderate size HII region. (The 15$'$
extended structure around the bi-annular source shows a
non-thermal spectrum and from it's emission measure at
8.35 GHz, the estimated optical depth due to free-free
absorption at 330 MHz is found to be negligible.)

(iii) The bi-annular structure itself is unlikely to
have any associated HII region. It was selected as a
candidate SNR due to it's low IR/Radio luminosity
\nocite{1989MNRAS.237..381B}({Broadbent}, {Osborne} \&  {Haslam} 1989). 
We estimate a flux density of
$\sim$ 400 MJy/sr from the surrounding region at 60
$\mu$ IRAS map. Extended HII region generally shows
a ratio of 700 for the IR/Radio luminosity
\nocite{1989MNRAS.237..381B}({Broadbent} {et~al.} 1989), 
which also indicate $\tau$ $\le$0.1.

Hence, the change in spectral index for this object
seems to be intrinsic in nature. Breaks in spectral
indices were earlier observed for a few Galactic SNRs
like Cygnus loop and HB9 \nocite{1974AJ.....79.1253D}({Denoyer} 1974),
S147 \nocite{1980A&A....92..225K}({Kundu} {et~al.} 1980), G33.2$-$0.6
\nocite{1982A&A...106..314R}({Reich} 1982), 3C58
\nocite{1994ApJS...90..817G}({Green} 1994) G76.9+1.0
\nocite{1997A&AS..123..199L}({Landecker} {et~al.} 1997).
However, only S147, G33.2$-$0.6 and G76.9+1.0 show break at
metre wavelengths.  From its irregular morphology and the
spectral break, we suggest that G357.1$-$0.2 may
be a Crab like SNR. The pulsar B1739$-$030 is located 2$'$
away, but, there is no indication that it is associated with
this object.

It is also possible that G357.1$-$0.2 is associated with the
large suspected SNR which we have named as G356.8$-$0.0.
G357.1$-$0.2 shows HI absorption by the 3 kpc arm.  This
implies a distance of more than 6 kpc from us for this
object. The distance of the large SNR G356.8$-$0.0 could
also be similar ($\sim$ 4.4 kpc from $\Sigma$-D relation).
However, the large SNR apparently being a shell type remnant
(expected spectral index $~\sim-$0.5), an association with
an object of different spectral type seems unlikely. We
suggest that G357.1-0.2 may be a crab like SNR, where the
energy transfer from the pulsar has stopped in the past,
giving rise to a steeper spectral-index at the higher
radio-frequency.  The estimated parameters for G357.1$-$0.2
are summarised in Table 3.



\begin{table*}
\begin{minipage}{150mm}
\caption{Compact sources in the field centred on G3.2$-$1.0}
\begin{tabular}{c c c c c c c c c}
\hline
  Name     & RA &  DEC  &     Peak & Total (0.33 GHz) & Total (1.4 GHz) &  Distance & Comment \\
           & (J2000) & (J2000) & (Jy/Beam) &  (Jy) & (Jy)    & from centre &        \\
              &         & &        &               &        & ($'$)     &          \\
\hline
$2.442-0.701$ & 17 54 02.0 & -27 12 02 &  0.030 &  0.056 & 0.003   & 50.9 &  \\ 
$2.546-0.917$ & 17 55 06.6 & -27 13 13 &  0.056 &  0.102 & 0.0137  & 41.6 &         \\
$2.548-1.309$ & 17 56 38.5 & -27 24 56 &  0.055 &  0.120 & 0.038   & 45.1 & Double   \\
$2.596-1.315$ & 17 56 46.3 & -27 22 38 &  0.067 &  0.095 & 0.019   & 42.7 &         \\
$2.596-1.086$ & 17 55 52.6 & -27 15 45 &  0.236 &  0.405 & 0.120   & 38.6 &  TX=0.397 \\
             & &        & &        &               &         &                    \\
$2.603-0.774$ & 17 54 40.9 & -27 05 56 &  0.026 &  0.032 & 0.0014  & 40.3 &  Pulsar ? \\
$2.749-1.518$ & 17 57 54.5 & -27 20 46 &  0.025 &  0.023 & 0.011   & 42.5 &         \\
$2.773-1.274$ & 17 57 00.7 & -27 12 15 &  0.035 &  0.053 & 0.0086  & 32.2 &         \\
$2.798-0.596$ & 17 54 26.1 & -26 50 26 &  0.046 &  0.068 & 0.012   & 35.8 &         \\
$2.816-1.091$ & 17 56 23.8 & -27 04 29 &  0.030 &  0.060 & 0.010   & 25.7 &         \\
             & &        & &        &               &         &                      \\ 
$2.833-1.218$ & 17 56 55.8 & -27 07 26 &  0.057 &  0.070 & 0.021   & 27.4 &         \\
$2.838-0.708$ & 17 54 57.7 & -26 51 46 &  0.023 &  0.034 & 0.012   & 29.6 &         \\
$2.859-1.147$ & 17 56 42.6 & -27 03 55 &  0.051 &  0.135 & 0.038   & 24.1 &         \\
$2.893-0.884$ & 17 55 45.9 & -26 54 15 &  0.019 &  0.021 & 0.007   & 21.7 &         \\
$2.924-1.304$ & 17 57 28.1 & -27 05 18 &  0.009 &  0.010 & 0.0014  & 26.0 &    \\ 
             & &        & &        &               &         &                     \\ 
$2.937-0.352$ & 17 53 48.5 & -26 35 49 &  0.033 &  0.059 & 0.008  & 42.9 &         \\
$2.961-0.656$ & 17 55 02.2 & -26 43 48 &  0.093 &  0.203 & 0.030  & 26.5 &   Double \\ 
$2.988-1.304$ & 17 57 36.7 & -27 01 57 &  0.012 &  0.020 & $<$0.001 & 23.4 &      \\ 
$3.016-0.372$ & 17 54 03.8 & -26 32 22 &  0.056 &  0.064 & 0.011  & 40.0 &         \\
$3.022-1.584$ & 17 58 46.6 & -27 08 34 &  0.019 &  0.032 & 0.014  & 37.1 &        \\
             & &        & &        &               &         &              \\ 
$3.026-1.186$ & 17 57 14.2 & -26 56 25 &  0.010 &  0.018 & 0.0016 & 16.7 &      \\ 
$3.044-1.136$ & 17 57 05.2 & -26 53 59 &  0.007 &  0.005 & 0.004  & 14.0 &        \\
$3.110-1.189$ & 17 57 26.4 & -26 52 11 &  0.018 &  0.048 & 0.007  & 13.5 &  Double \\
$3.136-0.775$ & 17 55 53.5 & -26 38 22 &  0.033 &  0.043 & 0.014  & 14.8 &  NV=0.004 \\
$3.157-0.556$ & 17 55 05.6 & -26 30 39 &  0.136 &  0.157 & 0.026  & 27.1 &  NV=0.015  \\
             & &        & &        &               &         &              \\ 
$3.170-0.260$ & 17 53 59.1 & -26 20 58 &  0.065 &  0.096 & $<$0.003  & 44.7 &  Pulsar ? \\ 
$3.323-0.991$ & 17 57 08.8 & -26 35 10 &  0.030 &  0.028 & 0.004  & 5.43 &        \\
$3.347-0.324$ & 17 54 37.6 & -26 13 47 &  0.099 &  0.240 & 0.042  & 41.2 &  Double, NV=0.019 \\
$3.384-0.999$ & 17 57 19.2 & -26 32 13 &  0.016 &  0.018 & 0.003   & 8.93 &         \\
$3.434-1.077$ & 17 57 43.8 & -26 31 59 &  0.055 &  0.072 & 0.024   & 12.7 &        \\
             & &        & &        &               &         &                    \\ 
$3.505-1.116$ & 17 58 02.5 & -26 29 27 &  0.070 &  0.105 & 0.031   & 17.5 &        \\
$3.565-1.009$ & 17 57 45.7 & -26 23 08 &  0.010 &  0.013 & 0.002   & 19.7 &        \\
$3.572-1.551$ & 17 59 52.6 & -26 38 59 &  0.013 &  0.016 & $<$0.001  & 38.6 &  Pulsar ? \\ 
$3.595-1.566$ & 17 59 59.1 & -26 38 12 &  0.038 &  0.058 & 0.020   & 40.1 &         \\
$3.621-1.348$ & 17 59 12.0 & -26 30 22 &  0.015 &  0.019 & 0.048   & 31.1 &          \\
             & &        & &        &               &         &                       \\
$3.630-1.090$ & 17 58 13.1 & -26 22 10 &  0.178 &  0.210 & 0.054   & 24.2 &        \\
$3.663-0.320$ & 17 55 19.4 & -25 57 17 &  0.040 &  0.120 & 0.050   & 48.3 &  Double, NV=0.005 \\
$3.717-0.938$ & 17 57 49.4 & -26 13 06 &  0.013 &  0.010 & 0.002   & 29.1 &        \\
$3.721-0.841$ & 17 57 27.5 & -26 09 59 &  0.011 &  0.022 & $<$0.001  & 30.6 &       \\ 
$3.727-0.240$ & 17 55 09.6 & -25 51 34 &  0.029 &  0.039 & $<$0.0015 & 54.3 &  Pulsar ? \\ 
             & &        & &        &               &         &                       \\ 
$3.746-1.222$ & 17 58 59.2 & -26 20 05 &  0.091 &  0.174 & 0.058   & 33.3 &   Double  \\
$3.748-0.802$ & 17 57 22.1 & -26 07 23 &  0.018 &  0.026 & 0.003   & 33.0 &         \\
$3.889-1.122$ & 17 58 55.1 & -26 09 40 &  0.016 &  0.025 & 0.007   & 39.8 &         \\
$4.027-1.301$ & 17 59 54.9 & -26 07 49 &  0.026 &  0.027 & 0.012   & 50.7 &         \\
              & &        & &        &               &         &                        \\
\hline

\end{tabular}
\end{minipage}
\end{table*}


\begin{table*}
\begin{minipage}{150mm}
\caption{Compact sources in the field centred on G356.8$-$0.1}
\begin{tabular}{lc c c c c c c c c c rc}

%
\hline
 Name            &  RA      & DEC     &   Peak   & Total  &  Total &  Total  & Distance from & Comment \\
                 &(J2000)   & (J2000) &         & (0.33 GHz) & (1.4 GHz) & (5 GHz)   & Centre         &   \\
                 &          &         & (Jy/beam) & (Jy)  &  (Jy)     &  (Jy)     &  ($'$)   &     \\
\hline
$356.001+0.022$   & 17 35 40.4 & -32 18 54 & 0.606 &  1.31  &  0.397(NV)  &   0.114     & 48.5 & TX=1.01     \\
$356.018+0.428$   & 17 34 06.2 & -32 04 52 & 0.129 &  0.26  &  0.032   &   0.0062    & 56.5 & TX=0.286 \\
$356.379+0.004$   & 17 36 42.5 & -32 00 23 & 0.076 & 0.139  &   0.005  &   0.005     & 26.2 & Double \\
$356.394+0.338$   & 17 35 25.2 & -31 48 49 & 0.140 &  0.184 &   0.061  &   0.015     & 35.9 &        \\
$356.433-0.044$   & 17 37 02.2 & -31 59 12 & 0.033 &  0.055 &   $<$0.0015 &     ---    & 22.2 & \\
                 &            &        &          &            &         &                            \\
$356.463-0.014$   & 17 36 59.5 & -31 56 43 & 0.073 &  0.16  &    0.21(NV)  &    0.0083   & 20.7 &  Double \\
$356.580+0.567$   & 17 34 59.3 & -31 32 00 & 0.042 &  0.073 &   0.031  &      ---    & 42.0 &        \\
$356.597-0.739$   & 17 40 13.4 & -32 13 08 & 0.206 &  0.33  &   0.042  &      ---    & 40.2 &        \\
$356.655-0.152$   & 17 38 01.6 & -31 51 26 & 0.069 &  0.081 &   0.047  &    14.4   & 9.37 &        \\
$356.805-0.310$   & 17 39 02.0 & -31 48 53 & 0.035 &  0.065 &  $<$0.001(NV) &     ---      & 12.6 & \\ 
                 &            &        &           &                      &         &                \\
$356.830+0.360$   & 17 36 26.3 & -31 26 05 & 0.295 &  0.415 & 0.18         &      0.026  & 27.5 & TX=0.739 \\
$356.837+0.378$   & 17 36 23.0 & -31 25 09 & 0.046 &  0.067 &  0.0185  &     ---      & 28.7 & \\ 
$356.884+0.465$   & 17 36 09.6 & -31 19 58 & 0.084 &  0.102 & 0.028      &       ---    & 34.3 &        \\
$356.906+0.081$   & 17 37 44.0 & -31 31 14 & 1.837 &  3.75  &   2.24 (NV)   &   0.487  & 12.5 & TX=3.50 \\
$356.965-0.074$   & 17 38 29.7 & -31 33 12 & 0.044 &  0.043 &   0.002   &       ---    & 10.0 & Pulsar ? \\ 
                 &            &        &          &            &         &                          \\
$357.036-0.559$   & 17 40 36.2 & -31 45 06 & 0.098 &  0.140 &   0.043  &      ---     & 30.9 &       \\
$357.067-0.401$   & 17 40 03.0 & -31 38 29 & 0.033 &  0.042 &   $<$0.002  &      ---    & 24.1 & \\
$357.098-0.220$   & 17 39 24.5 & -31 31 07 & 0.069 &  0.060 &   $<$0.0015 &      ---     & 19.2 &  Pulsar \\
$357.134+0.316$   & 17 37 22.6 & -31 12 07 & 0.043 &  0.046 &   0.018  &     0.005   & 31.9 &       \\
$357.280+0.211$   & 17 38 09.4 & -31 08 06 & 0.206 &  0.301 &   0.154  &     0.02    & 34.2 &         \\
                 &            &        &          &            &         &                             \\
$357.437-0.519$   & 17 41 26.4 & -31 23 25 & 0.312 &  0.470 & 0.222  &      ---     & 45.6 & TX=0.369 \\
$357.452-0.597$   & 17 41 47.3 & -31 25 07 & 0.144 &  0.201 & 0.047  &      ---     & 49.0 &         \\
$357.809-0.299$   & 17 41 29.2 & -30 57 29 & 0.093 &  0.12  &  0.085   &     0.020   & 61.7 &         \\
                 &            &        &          &            &         &                              \\
\hline

\end{tabular}
\end{minipage}
\end{table*}

\section{Compact sources in the field of view:}

Our 330 MHz maps allow us to image compact sources in the
field of view.  The sources were searched from our high
resolution (beam-size $\approx15-20{''}$) maps of the two
fields. A total of 44 sources have been identified from the
field G3.2$-$1.0, and 23 sources from the field
G356.8$-$0.1. Due to twice better sensitivity in the map of
the first field, we could detect almost a factor of two
higher number of sources in the field G3.2$-$1.0. We have
presented the parameters for the sources in Table 4 for the
field G3.2$-$1.0 and in Table 5 for the field G356.8$-$0.1.
In Table 4, (Field centred at l=3.2, b=$-$1.0) column-1
shows the Galactic Co-ordinate of the object, columns 2 and
3 give the Right Ascension and Declination (J2000) of the
object respectively and columns 4 and 5 give the peak and
total flux density at 330 MHz.  Column 6 gives the total
flux density of the object at 1.4 GHz. Column-7 lists the
distance of the object from the pointing centre of the
fields and column-8 gives any comments about the source. In
the last column (Comments), 'TX' denotes the flux of the
object as measured by the Texas Survey at 365 MHz.
The flux density of these sources at 1.4 and 5 GHz
have been quoted from the VLA Galactic plane
survey \nocite{1994ApJS...91..347B,1992ApJS...80..211H,1990ApJS...74..181Z}({Becker} {et~al.} 1994; {Helfand} {et~al.} 1992; {Zoonematkermani} {et~al.} 1990),
along with the NVSS map at 1.4 GHz.  NVSS
observations being of much higher sensitivity than the
Galactic Plane survey, many sources from our map which
are absent in the Galactic Plane survey were found to
have counterpart in the NVSS map. Also, to estimate
flux density for the extended sources, we have preferred the NVSS
maps to minimise any zero-spacing problem.  The sources
for which the 1.4 GHz flux is quoted from the NVSS, an
entry of '(NV)' is given beside the numerical value of
the total flux density. For Table 5 (field centred at
l=356.8, b=$-$0.0), the columns 1 to 6 have the same
meaning as above. However, column-7 indicates the
source flux density at 5 GHz and column-8 indicates the
distance of the object from the pointing Centre.
Column-9 shows any comments about the objects. In
Table 5, the 1.4 GHz flux is estimated for most of the
objects from the NVSS map.  The sources for which the
1.4 GHz flux density have been quoted from the Galactic plane
survey, the corresponding NVSS flux has been given in
the last column (Comments) by 'NV=flux'.
The Galactic plane survey being of intrinsically higher
resolution, we have compared the source positions in
both of our 330 MHz maps with respect to the above
survey.  The typical positional error of the sources
were found to be $\approx$4-5$^{''}$, part of
which is caused by systematic offset for all the
sources.

The flux density quoted for the sources in our list can
be affected by two kinds of errors. The first is 
due to an error in the absolute flux density scale,
where, all the sources will be affected by this error.
We expect, the error in our flux density scale to be
within $\pm15\%$. The second kind of error in the
flux of the objects can be caused by the primary beam
pattern of the antennas. Though, our maps have been
corrected by multiplying the maps by the inverse of the primary
beam pattern, any error in pointing and in determining
the primary beam pattern will persist.
Only the sources which are located near the FWHM (88$'$ at
330 MHz) or beyond will get seriously affected by
this type of an error. In our list of sources, the
objects which are located close or beyond the FWHM of
the antennas can have an estimated flux density which
may be off even by 20$-$40\% of their absolute flux due
to this error.  Hence, the flux density for these
sources should be treated with caution.

Most of the sources in our list show a non-thermal
steep spectral index and is expected to be background
extragalactic sources. A few sources having non-thermal
spectra also have a double-lobed morphology. These
sources are almost certainly to be extragalactic. Four
sources from Field I (G2.603$-$0.774, G3.170$-$0.260, 
G3.572$-$1.551 and G3.727$-$0.240) (Table 4) and one
source from Field II (G356.965$-$0.074) (Table 5)
appears to be compact 
and have a spectral index steeper than $-$1.9. We
classify these sources as possible pulsar candidates.

\section{Conclusions} Our observations of four
suspected supernova remnants close to the Galactic
Centre region have resulted in: 

(a) Confirmation that G3.1$-$0.6 
\nocite{1994MNRAS.270..847G}({Gray} 1994b) is supernova
remnant. This was the largest suspected SNR (28$'\times
49'$) detected during the Molonglo Galactic Centre
survey. We estimate a steep spectral index of $\le$
$-$0.64 for this object. In the NVSS map, the narrow
filamentary part at the eastern region of the shell is
found to be polarised upto a level of 20\%.  This
indicate an ordered magnetic field in the region and
also implies a limit on the RM along the line-of-sight
medium to be $\le$ 1000 rad m$^{-2}$. The $\Sigma$-D
relation indicates a probable distance of 7.7 kpc from
us for this object.  The object has a two shell
morphology which we interpret as due to expansion of
the SNR into an anisotropic ISM.

(b) The detection of two curved extended structures in the
Field II, which appears to be part of a large shell like structure 
(size $\approx52' \times 72'$) centred at G356.8$-$0.0, and having a
total flux density of 30 Jy at 330 MHz. A comparison of our
map with single dish maps at 8.35 and 14.35 GHz constrains
it's spectral index to be steeper than $-$0.34. The shell
type morphology and a non-thermal spectral index suggest it
to be a supernova remnant. Observations at even lower
frequency is recommended to better constrain it's spectral
index.  Using  $\Sigma$-D relation, a probable distance of
4.4 kpc has been estimated for this object. It appears to
be a very old supernova remnant.

(c) We find that the three previously suspected supernova
remnants, G356.6+0.1, G356.3$-$0.3 and G357.1$-$0.2
\nocite{1994MNRAS.270..847G}({Gray} 1994b) are located on an
extended object, which appears to be part of a large shell
as described above. We suggest that G356.6+0.1 and
G356.3$-$0.3 are actually parts of this large shell which
were accidentally identified as separate objects in the MOST
survey.

(d) The peculiar morphology and the spectral index of
the object G357.1$-$0.2 suggests that it is a separate
supernova remnant along the same line of sight of the
large shell (G356.8$-$0.0). It appears to be a
plerionic type of supernova remnant with a spectral
break around 1 GHz and is located at a distance of more
than 6 kpc from us.

\section*{Acknowledgements}

We thank Andrew Gray and Sanjay Bhatnagar for useful
discussions. Andrew Gray provided us the MOST images of
the four suspected SNRs. We also thank Govind Swarup
and Jayaram Chengalur who read the manuscript of this
paper and made useful comments. Finally, we thank the
staff of the GMRT that made these observations
possible. GMRT is run by the National Centre for Radio
Astrophysics of the Tata Institute of Fundamental
Research.
\bibliography{}

\label{lastpage}
\end {document}